\documentclass[twocolumn,aps,prd,10pt,superscriptaddress,longbibliography]{revtex4-1}
\usepackage{amssymb,amsmath}
\usepackage{bm}
\usepackage{dcolumn}
\usepackage{glossaries}
\usepackage{graphicx}
\usepackage[version=4]{mhchem}
\usepackage[usenames,dvipsnames,svgnames]{xcolor}
\usepackage{siunitx}
\usepackage{hyperref}
\hypersetup{colorlinks=true, linkcolor=blue}
\usepackage{xr}
\usepackage{makecell}

\hypersetup{
pdfnewwindow=true,      
colorlinks=true,        
linkcolor=Blue,         
citecolor=Blue,         
filecolor=Blue,         
urlcolor=Blue           
}

\newglossaryentry{kappa}{
  name={\ensuremath{\kappa}},      
  description={thermal conductivity}, 
  first={thermal conductivity (\ensuremath{\kappa})} 
}
\newacronym{dft}{DFT}{density functional theory}
\newacronym{md}{MD}{molecular dynamics}
\newacronym{mlp}{MLP}{machine-learned potential}
\newacronym{mlip}{MLIP}{machine-learning interatomic potential}
\newacronym{ml}{ML}{machine learning}
\newacronym{nep}{NEP}{neuroevolution potential}
\newacronym{nn}{NN}{neural network}
\newacronym{snes}{SNES}{separable natural evolution strategy}
\newacronym{mbd}{MBD}{many-body dispersion}
\newacronym{rmse}{RMSE}{root mean square error}
\newacronym{hnemd}{HNEMD}{homogeneous non-equilibrium molecular dynamics}
\newacronym{nemd}{NEMD}{non-equilibrium molecular dynamics}
\newacronym{emd}{EMD}{equilibrium molecular dynamics}
\newacronym{shc}{SHC}{spectral heat current}
\newacronym{mfp}{MFP}{mean free path}
\newacronym{bte}{BTE}{Boltzmann transport equation}
\newacronym{ald}{ALD}{anharmonic lattice dynamics}
\newacronym{aimd}{AIMD}{\emph{ab initio} molecular dynamics}
\newacronym{ann}{ANN}{artificial neural network}
\newacronym{dp}{DP}{deep potential}
\newacronym{gap}{GAP}{Gaussian approximation potential}
\newacronym{hcacf}{HCACF}{heat current autocorrelation function}
\newacronym{nist}{NIST}{national institute of standards and technology}
\newacronym{nqe}{NQE}{nuclear quantum effect}
\newacronym{rdf}{RDF}{radial distribution function}
\newacronym{scan}{SCAN}{strongly constrained and appropriately normed}
\newacronym{vdos}{VDOS}{vibrational density of states}
\newacronym{adf}{ADF}{angular distribution function}
\newacronym{ace}{ACE}{atomic cluster expansion}
\newacronym{3d}{3D}{three dimensional}
\newacronym{2d}{2D}{two dimensional}
\newacronym{pdos}{PDOS}{phonon density of states} 
\newacronym{lpdos}{LPDOS}{local phonon density of states} 
\newacronym{tpdos}{TPDOS}{total phonon density of states} 
\newacronym{itc}{ITC}{interfacial thermal conductance}
\newacronym{cbn}{\textit{c}BN}{cubic boron nitride}
\newacronym{eels}{EELS}{electron energy-loss spectroscopy}
\newacronym{si}{SI}{Supporting Information}

\DeclareSIUnit\angstrom{\text{Å}}
\DeclareSIUnit{\atom}{atom}
\DeclareSIUnit{\step}{step}
\DeclareSIUnit{\atomstepsecond}{\atom\step\per\second}

\begin{document}

\title{Interface phonon modes governing the ideal limit of thermal transport across diamond/cubic boron nitride interfaces}

\author{Xiaonan Wang}
\thanks{These authors contributed equally to this work.}
\affiliation{School of Science, Harbin Institute of Technology, Shenzhen, 518055, P. R. China}

\author{Xin Wu}
\thanks{These authors contributed equally to this work.}
\affiliation{Institute of Industrial Science, The University of Tokyo, Tokyo 153-8505, Japan}

\author{Penghua Ying}
\email{hityingph@tauex.tau.ac.il}
\affiliation{Department of Physical Chemistry, School of Chemistry, Tel Aviv University, Tel Aviv, 6997801,Israel}

\author{Zheyong Fan}
\affiliation{College of Physical Science and Technology, Bohai University, Jinzhou 121013, P. R. China}
 
\author{Huarui Sun}
\email{huarui.sun@hit.edu.cn}
\affiliation{School of Science, Harbin Institute of Technology, Shenzhen, 518055, P. R. China}
\affiliation{Ministry of Industry and Information Technology Key Laboratory of Micro-Nano Optoelectronic Information System, Harbin Institute of Technology, Shenzhen, 518055, P. R. China}
\affiliation{Collaborative Innovation Center of Extreme Optics, Shanxi University, Taiyuan, 030006, P. R. China}

\begin{abstract}
Understanding the ideal limit of interfacial thermal conductance (ITC) across semiconductor heterointerfaces is crucial for optimizing heat dissipation in practical applications. By employing a highly accurate and efficient machine-learned potential trained herein, we perform extensive non-equilibrium molecular dynamics simulations to
investigate the ITC of diamond/cubic boron nitride ($c$BN) interfaces. The ideal diamond/$c$BN interface exhibits an unprecedented ITC of 11.0 $\pm$ 0.1 GW m$^{-2}$ K$^{-1}$, setting a new upper bound for heterostructure interfaces. This exceptional conductance originates from extended phonon modes due to acoustic matching and localized C-atom modes that propagate through B-C bonds. However, atomic diffusion across the ideal interface creates mixing layers that disrupt these characteristic phonon modes, substantially suppressing the thermal transport from its ideal limit. Our findings reveal how interface phonon modes govern thermal transport across diamond/$c$BN interfaces, providing insights for thermal management in semiconductor devices.

\end{abstract}
\maketitle
\raggedbottom

\section{Introduction}

The advancement of semiconductor materials has ushered in a new era of micro/nano electronic devices. However, as device dimensions continue to shrink following Moore’s law, efficient heat dissipation has emerged as a critical technological challenge, particularly under peak operating conditions \cite{cheng2024ultra}. To optimize performance, modern microelectronic devices often integrate two or even more materials to leverage their complementary advantages, as seen in AlGaN/GaN high-electron mobility transistors \cite{zhu2019piezotronic}, diamond-based semiconductor (e.g., Ga$_2$O$_3$, SiC) radiofrequency devices \cite{xu2024first,cheng2020integration,ji2024interfacial}, Al/GaN deep-ultraviolet photoelectric applications \cite{tsao2018ultrawide}, and diamond/\gls{cbn} electronic devices \cite{bello2002diamond}. In these heterostructures, \gls{itc} plays a pivotal role in thermal management, as interfaces often serve as bottlenecks for thermal transport. Understanding the upper limit of \gls{itc} and underlying phonon transport mechanisms in heterostructures is thus crucial for optimizing semiconductor performance. 

Among various heterointerfaces, diamond/\gls{cbn} interfaces are particularly promising, as both crystalline diamond and \gls{cbn} are superhard materials with exceptionally high \gls{kappa} \cite{zhu2021thermal, 2020chenultrahigh}. Furthermore, their minimal lattice mismatch enables the formation of an atomically flat interface, making them ideal candidates for maximizing \gls{itc}. Experimental fabrication of high-quality, flat diamond/\gls{cbn} interfaces has been demonstrated \cite{chen2015misfit}, providing an ideal model system for investigating the theoretical upper limit of \gls{itc}. Recent advancements in \gls{eels} have enabled the probing of nanoscale interfacial phonon dispersions and phonon modes across the diamond/\gls{cbn} interfaces \cite{qi2021measuring,kikkawa2021nanometric}, providing direct evidence for the existence of specific phonon modes \cite{gordiz2016phonon}. However, directly quantifying the contributions of interface phonon modes to thermal conductance in experiments remains challenging due to the nature of buried interfaces. Furthermore, roughness and atomic interdiffusion are prevalent in realistic interfaces \cite{bello2005deposition, chen2021experimental}, necessitating an understanding of how interface phonon modes evolve from smooth to rough interfaces and their effects on \gls{itc} for advanced semiconductor thermal management.

Besides experimental efforts, several computational approaches have attempted to predict \gls{itc} values for diamond/\gls{cbn} interfaces. \Gls{md} simulations \cite{qi2021measuring,li2024molecular} based on Tersoff potential \cite{kinaci2012thermal} have been conducted, yet this empirical potential lacks parametrization for diamond/\gls{cbn} interfaces and fails to accurately capture interface phonon dispersions, severely limiting the reliability of \gls{itc} predictions. Alternatively, Monte Carlo simulations incorporates phonon properties from \gls{dft} calculations \cite{xu2020high} have been used to model steady-state thermal transport across the interface. However, these simulations rely on acoustic and diffusion mismatch models, which assume purely elastic phonon transport through ballistic or diffusive mechanisms, thereby neglecting the contributions of inelastic phonon transport—an essential factor in real interfaces with complex atomic configurations \cite{chen2022interfacial,li2022inelastic}. To address these challenges, \glspl{mlp} have emerged as a promising alternative \cite{deringer2019machine, dong2024molecular, ying2025advances}. By being trained on energy, atomic forces, and virial data from first principles calculations such as \gls{dft} \cite{behler2007generalized}, \glspl{mlp} enable \gls{md} simulations with near quantum mechanical accuracy while achieving computational speeds several orders of magnitude higher than \gls{aimd}. To date, \gls{mlp} driven \gls{nemd} simulations have been applied to investigate the  thermal conductance of semiconductor interfaces, including Ga$_2$O$_3$/diamond \cite{sun2024insight}, GaN/cubic boron arsenide \cite{wu2024deep} and Si/Ge \cite{chen2021experimental}, etc. 

\begin{figure*}[htb]
\begin{center}
\includegraphics[width=1.8\columnwidth]{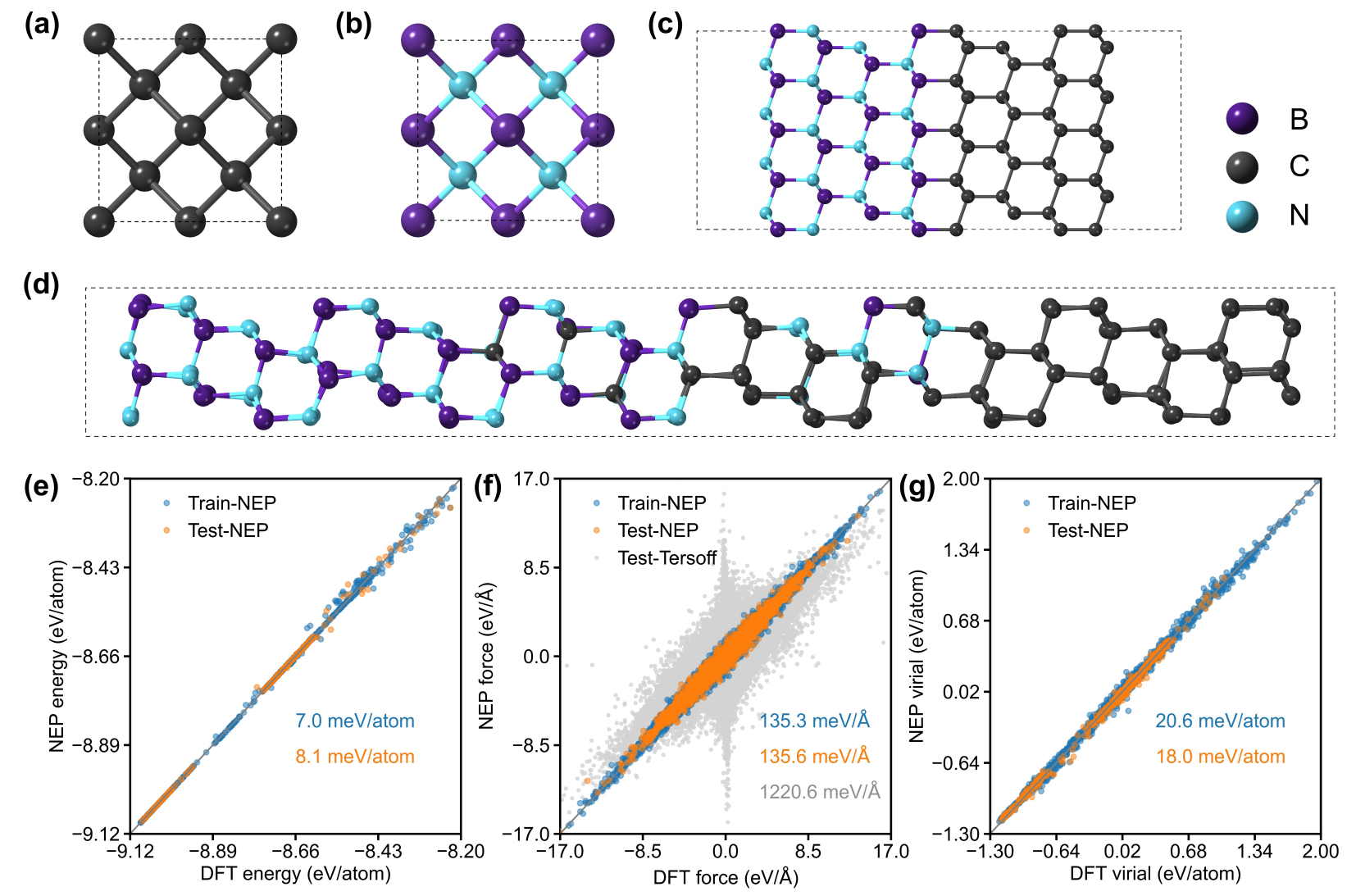}
\caption{Construction of the NEP model. (a)-(d) Reference structures used for training, including (a) diamond, (b) \gls{cbn}, and their heterostructures with (c) flat and (d) rough interfaces. (e)-(g) Comparison of (e) energy, (f) force, and (g) virial predicted by the \gls{nep} model against \gls{dft} reference data for both training and test datasets. In panel (f), atomic forces predicted by the Tersoff \cite{kinaci2012thermal} potential for the test dataset are also shown for comparison.}
\label{fig:nep}
\end{center}
\end{figure*}

In this study, we develop a unified machine-learned \gls{nep} \cite{fan2021neuroevolution, fan2022gpumd} model for pure diamond and \gls{cbn}, as well as diamond/\gls{cbn} heterostructures (see \autoref{fig:nep} (a-d)). We employ the \gls{nep} approach for its superior efficiency among existing \gls{mlp} frameworks. After demonstrating its accuracy and reliability, we apply the developed \gls{nep} to perform extensive \gls{nemd} simulations to study the thermal transport across diamond/\gls{cbn} interfaces. Our \gls{nemd} simulations predict an exceptionally high \gls{itc} for ideal flat diamond/\gls{cbn} interfaces, which surpasses all existing heterointerface \gls{itc} results reported in both experimental measurements and theoretical predictions. This remarkably high \gls{itc} stems from interface phonon modes, particularly extended and localized modes, that facilitate efficient phonon transport. We also demonstrate the atomic diffusion at rough interfaces suppresses these modes, leading to a significant reduction in \gls{itc}.

\section{Results and discussions}
\subsection{Validation of the NEP model}
\label{section:mlp}
To describe interatomic interactions across diamond/\gls{cbn} interfaces, we employed the third generation of \gls{nep} framework\cite{fan2022gpumd} to train a unified machine-learned \gls{nep} for diamond, \gls{cbn}, and their heterostructure (see \autoref{fig:nep}(a-d)). This is achieved by applying feedforward \gls{nn} together with \gls{snes} \cite{Schaul2011} to learn the energy, force, and virial of reference structures obtained from \gls{dft} calculations (see \autoref{section:dft} and \autoref{section:nep} for details). As shown in \autoref{fig:nep}(e-g), the developed \gls{nep} achieves high accuracy for both training and test datasets compared with \gls{dft} results, with \gls{rmse} values of energy, force, and virial less than \SI{8.2}{\meV\per\atom}, \SI{136}{\meV\per\angstrom}, and \SI{21}{\meV\per\atom}, respectively. Notably, for the atomic forces in the test dataset, the \gls{rmse} of \gls{nep} is approximately an order of magnitude lower than that of the Tersoff potential, which was used to drive \gls{md} simulations for predicting the \gls{itc} of diamond/\gls{cbn} interfaces in previous studies \cite{qi2021measuring, li2024molecular}.  

\begin{figure}[htbp]
 \begin{center}
 \includegraphics[width=0.48\textwidth]{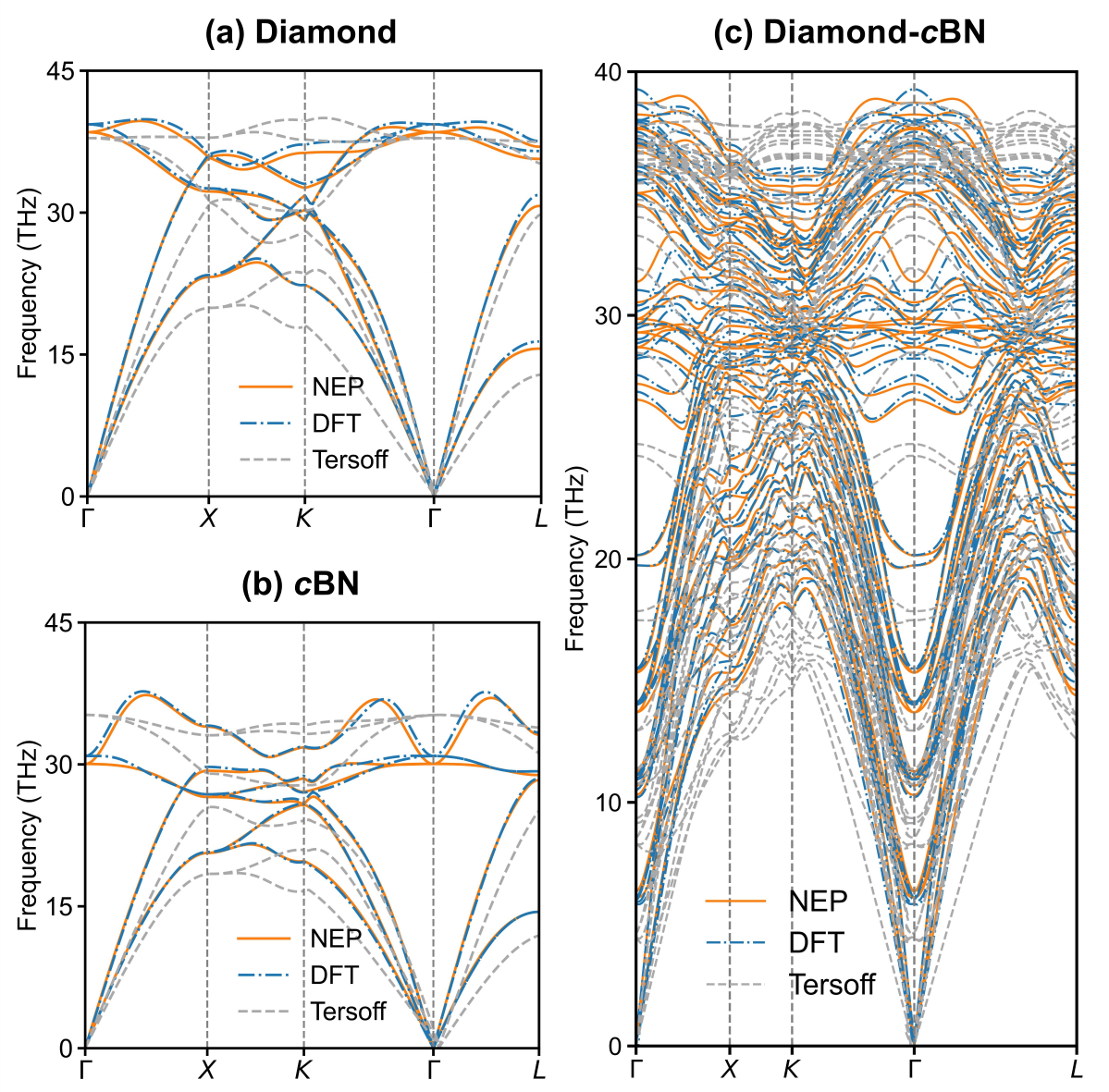}
 \caption{Phonon dispersion bands of (a) diamond, (b) \gls{cbn}, and (c) diamond/\gls{cbn} heterostructure predicted using \gls{dft}, \gls{nep}, and the Tersoff \cite{kinaci2012thermal} approach.}
 \label{fig:phonon}
 \end{center}
 \end{figure}

In \autoref{fig:phonon}, we further validate the accuracy of the \gls{nep} model and the Tersoff potential in describing the phonon dispersions of diamond, \gls{cbn}, and their ideal interface (see \autoref{section:dft} for details). The unified \gls{nep} accurately reproduces the phonon dispersion not only for bulk diamond and \gls{cbn} (see \autoref{fig:phonon}(a-b)) but also for their heterostructures (see \autoref{fig:phonon}(c)). In contrast, the Tersoff potential exhibits apparent deviations, particularly showing a strong softening effect on the acoustic branches.

Before performing our \gls{itc} calculations for the diamond/\gls{cbn} heterostructure, we first applied the developed \gls{nep} to conduct \gls{hnemd} simulations and predict the \gls{kappa} of the bulk counterparts at \SI{300}{\kelvin} (see \gls{si} section S1 for details). The predicted $\kappa$ for diamond and \gls{cbn} are \SI{2215 \pm 70}{\watt\per\meter\per\kelvin} and \SI{1223 \pm 8}{\watt\per\meter\per\kelvin}, respectively. In addition to the \gls{nep}-\gls{hnemd} approach, we also employed the \gls{bte} method to predict the $\kappa$ for both diamond and \gls{cbn}. And in the \gls{dft}-\gls{bte} approach, harmonic and thirdorder anharmonic force constants considered here were derived from \gls{dft} calculations performed at the same accuracy level as those used for \gls{nep} training (see \autoref{section:dft} for details). In \autoref{table:comparisons}, we note that our \gls{nep}-\gls{md} prediction is lower than the \gls{dft}-\gls{bte} predictions, which can be potentially caused by the neglection of higher order phonon anharmonicity or phonon coherence effects in current \gls{bte} predictions.

We also compare our \gls{nep} and \gls{dft}-\gls{bte} predictions with reported measurement results, as well as previous \gls{md} predictions based on Tersoff potential in \autoref{table:comparisons}. For experimental data, only the maximum \cite{zhu2021thermal,2020chenultrahigh} and minimum \cite{slack1973nonmetallic,morelli2006high,slack1973nonmetallic} reported $\kappa$ values are listed to reflect variations due to sample quality, size and inevitable defects or impurities. In terms of the $\kappa$ of two bulks, our \gls{nep} prediction falls within the range of corresponding experimental measurements. In contrast, the \gls{md} simulations based on Tersoff potential greatly underestimated the $\kappa$ of two bulks, possibly due to their softening effect on phonon dispersions (see \autoref{fig:phonon}). The good agreement between our \gls{nep} predictions and experimental measurements, combined with validations of atomic forces and interfacial phonon dispersions, confirms the reliability of \gls{nep} in modeling thermal transport across diamond/\gls{cbn} interface.

\begin{table}[htb]
\setlength{\abovecaptionskip}{1cm}  
\caption{Comparison of the \gls{kappa} of diamond and \gls{cbn} at \SI{300}{\kelvin} predicted by different approaches. All values are given in units of \SI{}{\watt\per\meter\per\kelvin}, with values in parentheses representing the standard error.}
\centering
\begin{tabular}{llll}
\hline
Method & Diamond & \gls{cbn} \\
\hline
Our \gls{nep}-\gls{md} & 2215(70) & 1223(8)\\
Our \gls{dft}-\gls{bte} & 3362 & 1721\\
Tersoff-\gls{md}   & 1859\textsuperscript{\cite{li2024molecular}}   & 763\textsuperscript{\cite{li2024molecular}}\\
Experiment & 2400(50),\textsuperscript{\cite{zhu2021thermal}} 2000\textsuperscript{\cite{slack1973nonmetallic}}   & 1600(170),\textsuperscript{\cite{2020chenultrahigh}} 760\textsuperscript{\cite{morelli2006high}}\\
\hline
\label{table:comparisons}
\end{tabular}
\end{table}

\subsection{Ideal limit of interfacial thermal conductance}
\label{section:ideal_interface}
We then employ the trained \gls{nep} to perform \gls{nemd} simulations for diamond/\gls{cbn} heterostructure (see \autoref{section:nemd} for details). We first investigate the thermal transport across the ideal atomically flat interface of the diamond/\gls{cbn} heterostructure. \autoref{fig:nemd}(a) illustrates the model setup for \gls{nemd} simulations, where heat flux is transported from the left diamond side, across the central interface, to the right \gls{cbn} side. To calculate the temperature profile (see \autoref{fig:nemd}(b)), we divide the whole system into different groups along thermal transport direction, with each group representing a layer of diamond on the left side of the interface (labeled as $L1$, $L2$, $L3$, …) and a corresponding layer of \gls{cbn} on the right ($R1$, $R2$, $R3$, …). The interface temperature difference $\Delta T$ used for predicting \gls{itc} in the \gls{nemd} simulations (see \autoref{equation:G}) is consistently defined as the temperature difference between $L6$ and $R6$, rather than $L1$ and $R1$, in all cases. This definition of $\Delta T$ ensures a fair comparison between the \gls{itc} of the ideal interface and that of rough interfaces, where atomic diffusion is confined between $L6$ and $R6$. A detailed discussion on rough interfaces is provided in \autoref{section:rough_interface}. As shown in the inset of \autoref{fig:nemd}(b), the equal absolute slopes of energy injection from heat source and extraction from heat sink confirm energy conservation and validate the steady-state regime. Within this steady-state window, temperature profile analysis reveals a distinct temperature discontinuity at the interface, attributable to interfacial thermal resistance. 

\begin{figure}[htbp]
\begin{center}
\includegraphics[width=0.45\textwidth]{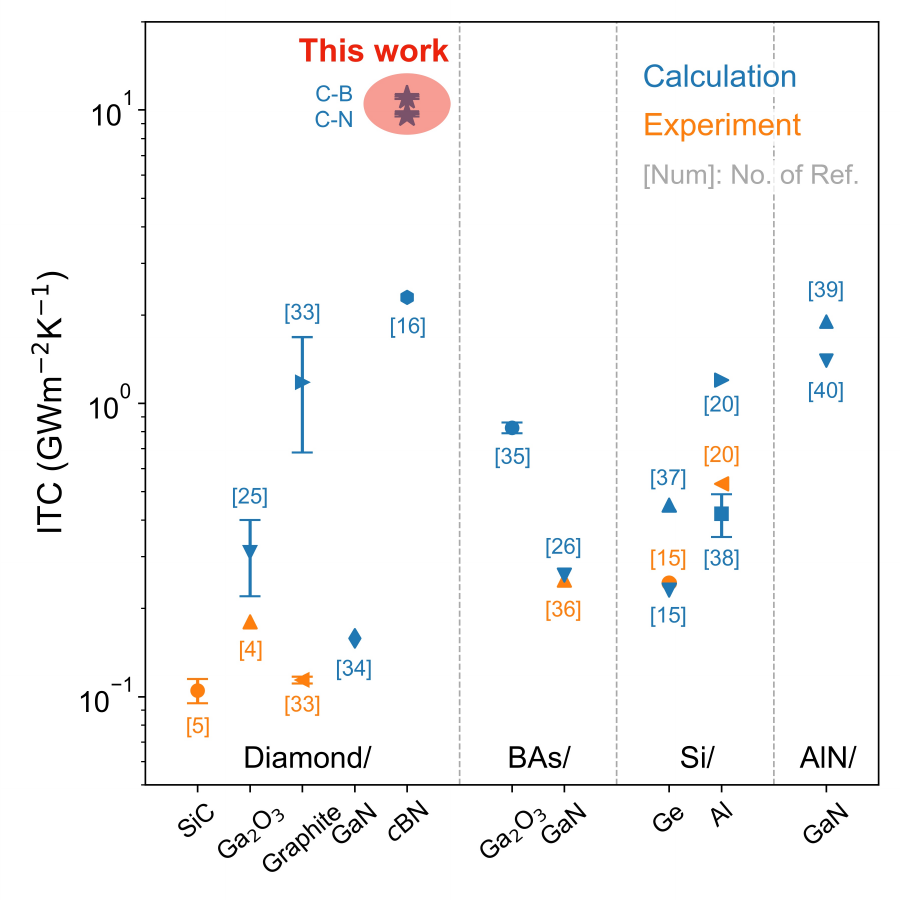}
\caption{The \glspl{itc} across diverse  representative heterostructures, comparing experimental measurements and theoretical predictions. Experimental data (orange) were obtained via time-domain thermoreflectance. Theoretical results(blue) include predictions from \gls{md} simulations using empirical potentials and \glspl{mlp}. Heterostructures shown encompass diamond-SiC\cite{ji2024interfacial}/Ga$_2$O$_3$\cite{cheng2020integration,sun2024insight}/graphite\cite{qiu2025ultra}/GaN\cite{wu2024significantly}/\gls{cbn}\cite{li2024molecular}, BAs-Ga$_2$O$_3$\cite{zhou2025ultra}/GaN \cite{wu2024deep,kang2021integration}, Si-Ge\cite{chen2021experimental,feng2019unexpected}/Al \cite{li2022inelastic,rustam2022optimization}, and AlN-GaN \cite{xue2025optimizing,koh2009heat}interfaces.}
\label{fig:comparison}
\end{center}
\end{figure}

To form a heterostructure through ideal interfaces, diamond can establish covalent bonds with \gls{cbn} through either C-B or C-N bonded pairs. Based on five independent \gls{nemd} simulations, we predict an \gls{itc} of \SI{11.0 \pm 0.1}{\giga\watt\meter^{-2}\kelvin^{-1}} for the C-B binding interface and \SI{9.6 \pm 0.09}{\giga\watt\meter^{-2}\kelvin^{-1}} for the C-N binding interface. As shown in \autoref{fig:comparison}, the predicted \gls{itc} values for diamond/\gls{cbn} are unprecedentedly high compared with existing theoretical and experimental results across various heterostructure interfaces. This significantly exceeds previous findings and appears to represent the ideal limit of \gls{itc} for solid heterointerfaces. As the \gls{itc} values for diamond/\gls{cbn} with C-B and C-N bonded interfaces are very close, in the following discussion we focuse on the C-B bonded case, which exhibits the highest \gls{itc} observed, while the corresponding results for the C-N bonded interface are presented in \gls{si} Section S2.

\subsection{Interface phonon modes}
\label{section:interfacial_modes}
To understand the mechanism underlying the observed upper bound of \gls{itc}, we perform phonon analyses in this section. Specifically, we elucidate the role of interfacial phonon modes in governing phonon transport across the ideal diamond/\gls{cbn} interface.

\begin{figure}[htb]
\begin{center}
\includegraphics[width=1\columnwidth]{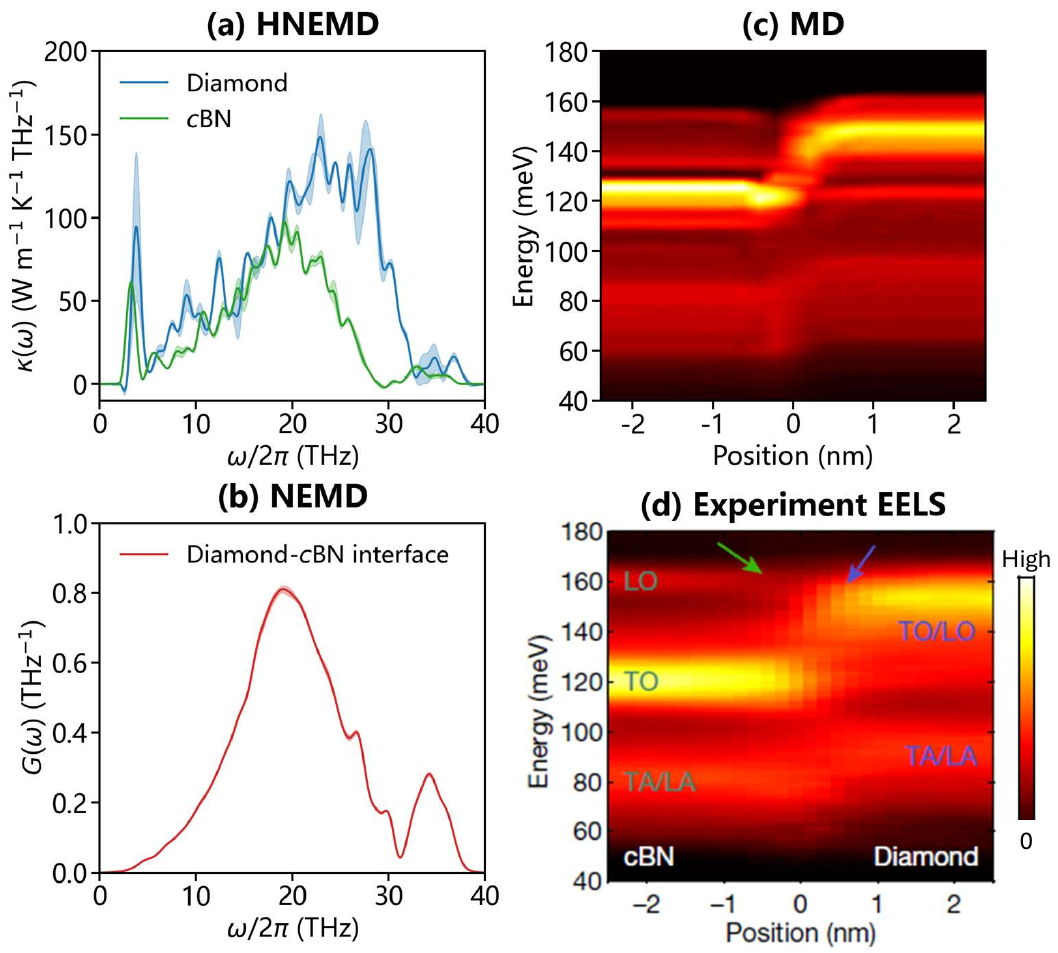}
\caption{(a) The spectrally decomposed $\kappa$ of diamond and \gls{cbn} obtained from \gls{hnemd} simulations. (b) Spectral decomposed thermal conductance of the diamond and \gls{cbn} ideal interface obtained from \gls{nemd} simulations. (c) Calculated normalized \gls{pdos} projected onto atom layers near the interface. (d) The measured \gls{eels} line profile across the interface. Reproduced with permission from ref \cite{qi2021measuring}. Copyright 2021, Springer Nature.}
\label{fig:shc}
\end{center}
\end{figure}

Based on the \gls{shc} analysis (see \autoref{section:shc} for details) \cite{fan2017thermal}, \autoref{fig:shc}(a) presents the phonon frequency dependent $\kappa$ of bulk diamond and \gls{cbn} from \gls{hnemd} simulations, and \autoref{fig:shc}(b) shows the \gls{itc} of their ideal interface as a function of phonon frequency from \gls{nemd} simulations. The \gls{itc} profile exhibits two prominent peaks at around the \SI{20} {\tera\hertz} and \SI{34} {\tera\hertz} (see \autoref{fig:shc}(b)), which identifies the dominant phonon modes contributing to interfacial thermal transport. While the \SI{20} {\tera\hertz} peak corresponds to the maximum overlap of the spectrally decomposed $\kappa$ of two bulk counterparts (see \autoref{fig:shc}(a)), the \SI{34}{\tera\hertz} peak emerges without an apparent bulk contribution. This unexpected peak demonstrates that interfacial phonon transport cannot be explained solely by bulk phonon properties, suggesting the presence of additional phonon contributions originating from the interface. As shown in the \gls{pdos} distribution near the interface (see \autoref{fig:shc}(c)), a frequency shift is observed as the region approaches the interface, resulting in the emergence of the \SI{34}{\tera\hertz} phonon mode (corresponding to approximately 140 meV). This observation of distinct interface phonon modes is consistent with previous \gls{eels} measurements (see \autoref{fig:shc}(d)) \cite{qi2021measuring}. 

\begin{figure}[htb]
\begin{center}
\includegraphics[width=1\columnwidth]{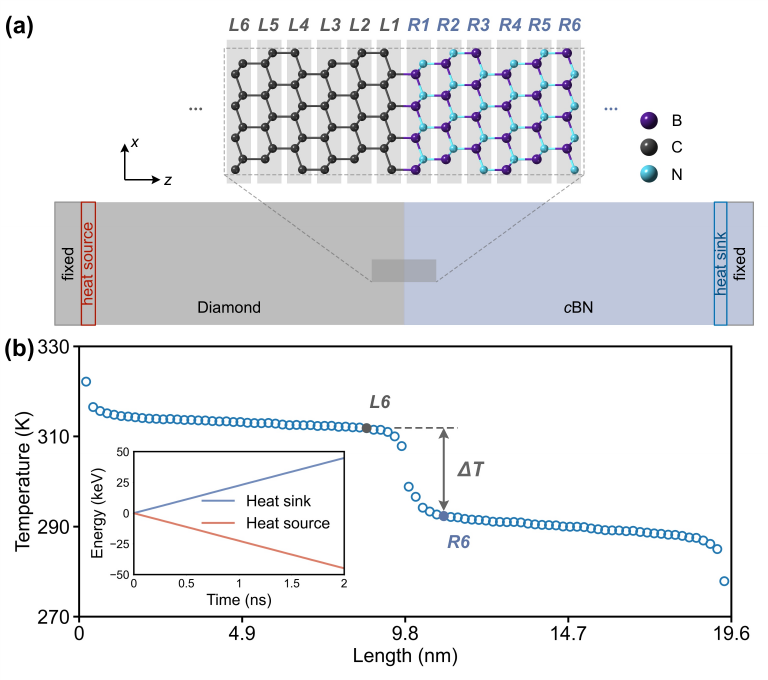}
\caption{(a) Schematic diagram of the \gls{nemd} setup. (b) The temperature profile of the diamond/\gls{cbn} ideal interface obtained from \gls{nemd} simulation at \SI{300} {\kelvin}. The insert shows the energy accumulated of the thermostats coupled with heat source or heat sink.}
\label{fig:nemd}
\end{center}
\end{figure}

\begin{figure*}[htb]
\begin{center}
\includegraphics[width=1.8\columnwidth]{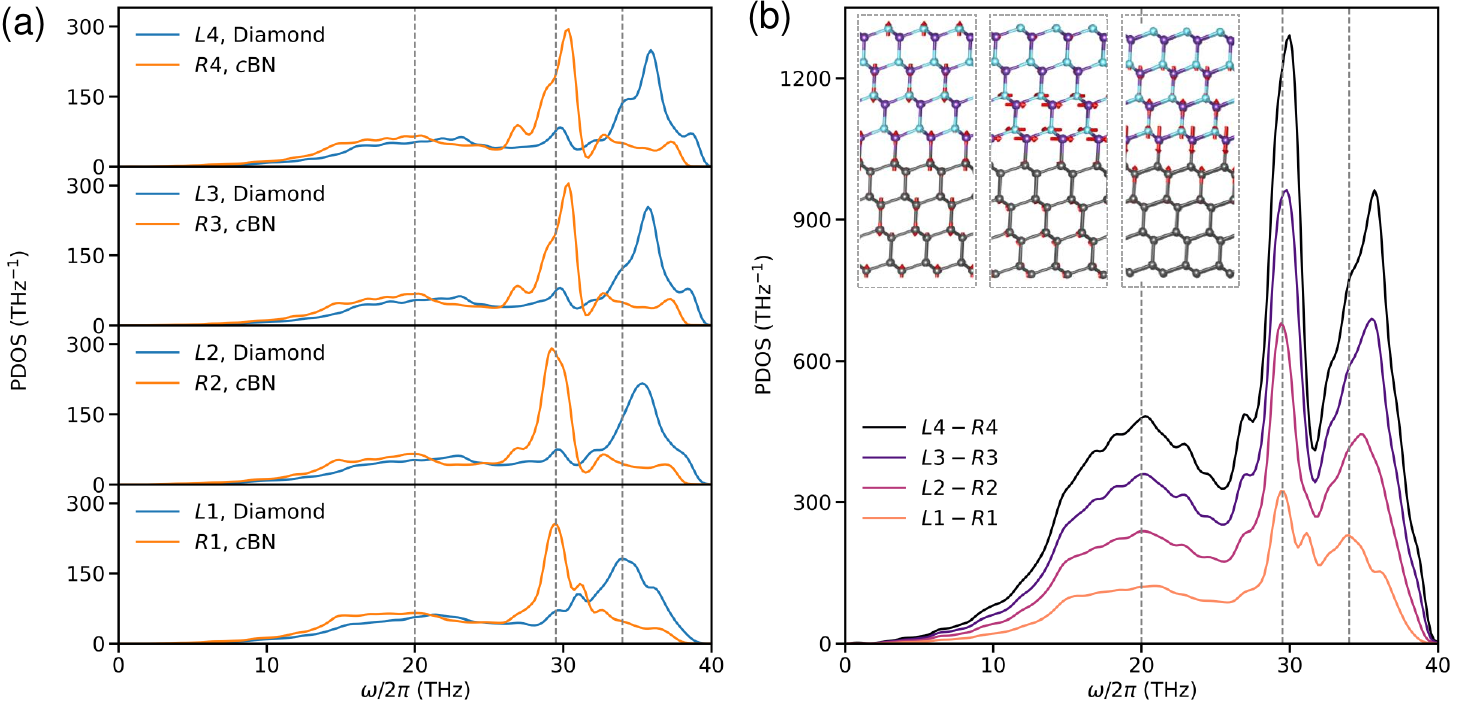}
\caption{(a) The local \gls{pdos} of each representative layer in the ideal interface structure. (b) The  total \gls{pdos} of multiple layers near the interface. The annotation ``$Li$-$Ri$'' in the figure represents the total  \gls{pdos}  of the central region from $Li$ to $Ri$ layers, which $i$ denotes the group label index of diamond on the left side and \gls{cbn} on the right side (see the inset in \autoref{fig:nemd}(a)). The inset in panel (b) shows the visualization of phonon eigenvectors at representative frequencies.}
\label{fig:pdos}
\end{center}
\end{figure*}

To reveal the formation mechanism of interface phonon modes, \autoref{fig:pdos}(a) analyzes the local \gls{pdos} evolutions for representative atomic layers. The diamond layer adjacent to the interface ($L1$ group in \autoref{fig:nemd}(a)) exhibits a distinctive spectral signature with a primary peak at approximately \SI{34}{\tera\hertz} and a secondary peak at \SI{31}{\tera\hertz}, corresponding to the B-C bond vibration detectable at \SI{1430}{cm^{-1}} via Fourier transforms infrared spectroscopy\cite{romanos2013infrared}. Progressing to the $L2$ group of diamond,a blue shift in the primary peak is observed, while the secondary peak vanishes entirely. From the $L3$ group onward, the \gls{pdos} gradually stabilizes and becomes indistinguishable from layers farther from the interface (e.g., the $L4$ group), indicating a transition to bulk phonon characteristics. Similar behavior is observed on the right \gls{cbn} side. The layer-resolved \gls{pdos} analysis clearly demonstrates that interfacial phonon vibration modes differ substantially from those in the bulk regions. 

For more comprehensive insight, we calculated the aggregate \gls{pdos} for multiple near-interface layers from left diamond to right \gls{cbn} sides, as shown in \autoref{fig:pdos}(b). As the interface region expands from the $L1$-$R1$ region to $L4$-$R4$ region, the relative contributions of phonon modes around \SI{34.5}{\tera\hertz} and \SI{29.5}{\tera\hertz} diminishes due to the incorporation of vibration modes from more distant layers, resulting in a blue shift of these peaks. Beyond approximately eight layers from the interface ($L4$-$R4$), the \gls{pdos} peak positions stabilize due to the predominance of bulk-like phonon modes. Notably, the peak at approximately \SI{20}{\tera\hertz} remains consistently positioned across all regions, indicating its presence in both interface and bulk environments. The inset of \autoref{fig:pdos}(b), we further visualizes the eigenmodes and analyzed atomic vibration amplitude eigenvectors of these three phonon modes (\SI{20}{\tera\hertz}, \SI{29.5}{\tera\hertz}, \SI{34.5}{\tera\hertz}) across eight layers in the interface region ($L4$-$R4$, spanning approximately \SI{1.7}{\nano\meter}). The \SI{20}{\tera\hertz} mode displays significant vibration amplitude throughout the entire interface region. The \SI{34.5}{\tera\hertz} mode shows substantial amplitude exclusively at the immediate interface, and is confined primarily to two atomic layers. Similarly, the \SI{29.5}{\tera\hertz} mode is localized at the interface but stems predominantly from B and N atomic vibrations at the boundary, which explains its negligible contribution to \gls{itc} (\autoref{fig:shc}(b)). Our findings provide conclusive evidence for the spatial evolution behavior of the interfacial localized phonon modes and their respective contributions to \gls{itc}, complementing the experimental observation \cite{qi2021measuring}. These interfacial modes provide an additional pathway for phonon transport across the diamond/\gls{cbn} heterointerface, and they are fundamentally distinct from those in the respective bulk materials.

The different contribution of each representative phonon mode to \gls{itc} are derived from the different origins of their vibrations. Phonon modes around \SI{34} {\tera\hertz} are highly localized at the interface, called the localized modes. These localized vibrational modes, primarily induced by interfacial C atoms (see \autoref{fig:pdos}(a)), are stabilized by strong C-C bonds. Importantly, the presence of the C-B bonds enable the extension of these vibrational modes to B atoms on the \gls{cbn} side. Further, the propagation direction of these phonons aligns with the thermal transport direction (see right inset in \autoref{fig:pdos}(b)), thereby contributing significantly to \gls{itc}. Another notable peak, around \SI{20} {\tera\hertz}, corresponds to acoustic phonon modes on both sides of the interface vibrating simultaneously (see left inset in \autoref{fig:pdos}(b)). These are delocalized across the entire heterostructure and thus are known as the extended modes. In contrast, the phonon mode near \SI{29.5} {\tera\hertz} involves horizontal vibrations at the interface that do not align with the effective heat transfer direction (see central inset in \autoref{fig:pdos}(b)), resulting a negligible contribution to \gls{itc}. Therefore, both extended and localized modes play crucial roles to interfacial heat transport, with the former contributing more significantly. Together, these modes result in the highest known \gls{itc} for a semiconductor heterostructure, enabled by the ideally flat atomic interface formed through strong bonded pair and minimal mass mismatch.

\subsection{Rough interface suppresses phonon transport}
\label{section:rough_interface}
\begin{figure}[htb]
\begin{center}
\includegraphics[width=1\columnwidth]{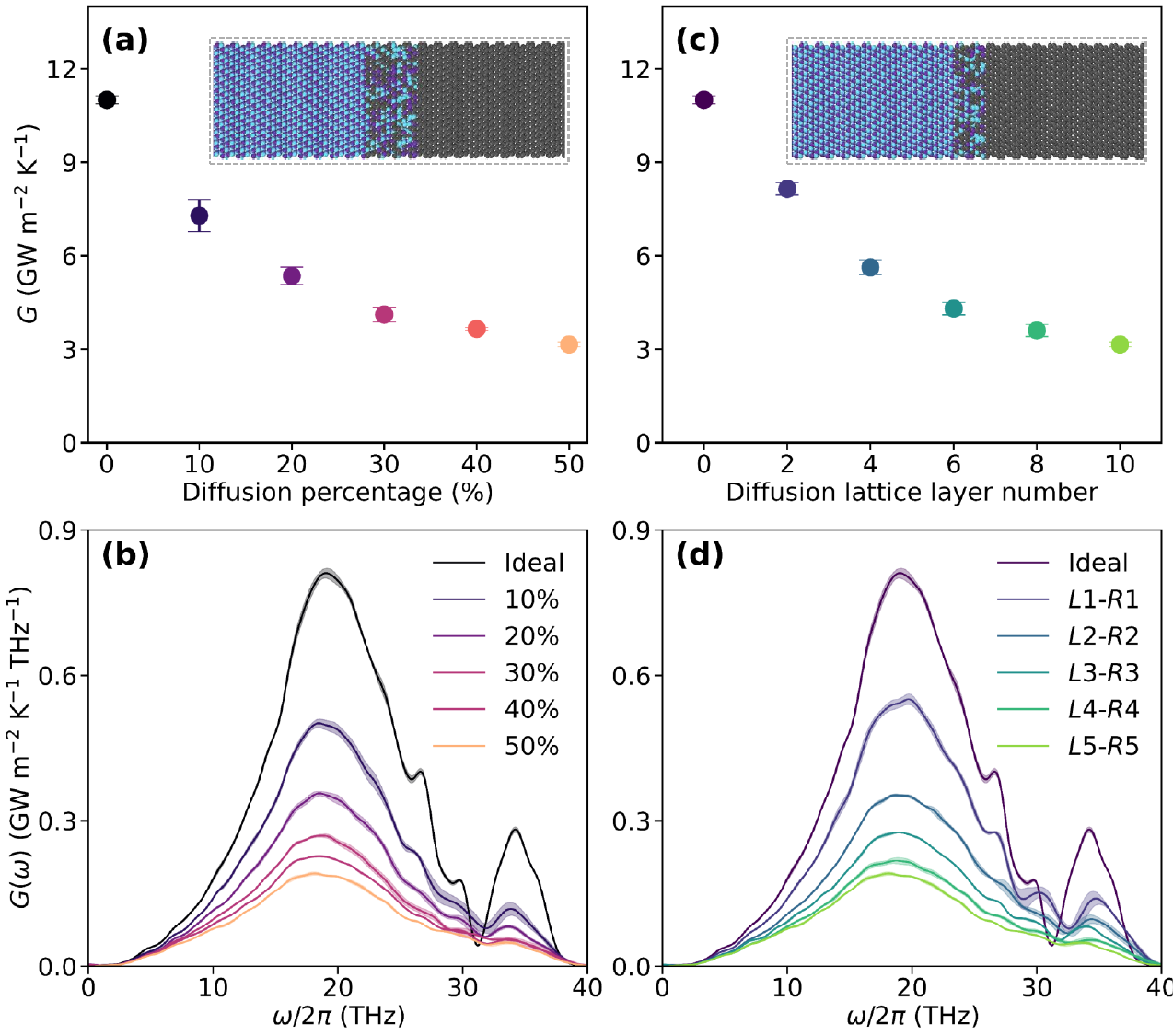}
\caption{Effect of atomic diffusion on the \gls{itc} of diamond/\gls{cbn} heterostructure. (a) \gls{itc} and (b) its spectral decomposition for 10-layer diffusion interfaces with varying diffusion ratios. (c) \gls{itc} and (d) its spectral decomposition for a fixed 50\% diffusion ratio with varying numbers of diffusion layers. The inset in (a) is an example of an interface with a 10-layer thick diffusion region with 50\% diffusion ratio, while the inset in (b) is an interface with a 6-layer thick diffusion region with 50\% diffusion ratio.}
\label{fig:rough}
\end{center}
\end{figure}

In this section, we further investigate the effect of interfacial atomic diffusion (i.e., roughness) on the \gls{itc} of diamond/\gls{cbn} heterostructures. For this purpose, we performed a series of ITC calculations on diamond/\gls{cbn} systems with rough interfaces. All systems involved use the same size as the ideal interfacial system, but randomly rearrange the atoms in the central region (2, 4, 6, and 10 layers, respectively) and achieve atomic mixing in different proportions (10\% - 50\%). Two cases are considered: (i) atomic diffusion confined within a fixed 10-layer region (from $L5$ to $R5$) with varying diffusion ratios (refer to the insert in \autoref{fig:rough}(a)), and (ii) atomic diffusion with a fixed 50\% mixing ratio but confined within different numbers of diffusion layers (refer to the insert in \autoref{fig:rough}(c)). To ensure the accuracy of the \gls{itc} results, we performed three independent \gls{nemd} simulations using distinct random atomic configurations for each rough interface. For both cases, it can be found that the atomic diffusion results in significant suppression of \gls{itc} (see \autoref{fig:rough}(a) and (c)). Compared to the ideal interface, the maximum 50\% atomic diffusion within in $L5$-$R5$ region results in approximately a 71\% reduction in \gls{itc}. It is worth noting that some previous studies have reported the opposite trend, where diffuse interfaces with appropriate atomic mixing, such as in Si/Ge \cite{chen2021experimental}, GaN/diamond \cite{wu2024significantly}, and Al/Si \cite{rustam2022optimization}, can enhance \gls{itc}.

The main reason for this inconsistency lies in the role of diffuse layers at the interface. Generally, for interfaces where the phonon modes differ significantly between the two bulk sides, diffuse layers can serve as a bridge to smooth the mismatch and facilitate phonon transport. In contrast, for interfaces where the phonon modes are already well matched, the introduction of diffuse layers disrupts this alignment and hinders thermal transport. In other words, whether the diffuse transition layers enhance or suppress phonon transport at the interface depends on the similarity of phonon modes across the interface. In the ideal flat interface of diamond/\gls{cbn} heterostructure, the enhanced \gls{itc} arises from the presence of well-matched phonon modes on both bulk sides of the interface. This can be attributed to two key factors: (i) the strong matching of acoustic phonon modes around \SI{20}{\tera\hertz} facilitates the formation of extended modes that contribute significantly to \gls{itc}, and (ii) the alignment of high-frequency optical phonon modes around \SI{34}{\tera\hertz} enables localized modes to also participate effectively in heat transfer. Given that diamond and \gls{cbn} share similar lattice structure and exhibit very similar acoustic phonon properties, the presence of atomic diffusion at the interface can significantly disrupt the overall vibrational coherence. This disruption weakens the contribution of phonon modes across the interface. As a result, the contribution of extended modes to \gls{itc} is notably reduced, as evidenced by the spectral decomposition results shown in \autoref{fig:rough}(b) and (d). Furthermore, as the interface becomes increasingly rough, phonon mode matching becomes more difficult, and the distinct vibrational features that enable high interfacial thermal transport in the ideal interface gradually vanish (see \gls{si} section S3).

\section{Conclusions}
\label{section:summary}

To probe the ideal upper limit of \gls{itc} for semiconductor heterostructures, we developed a unified \gls{mlp} for diamond, \gls{cbn}, and their heterointerfaces. Based on extensive NEP-driven \gls{nemd} simulations at near first-principles accuracy, the room temperature \gls{itc} of the ideal diamond/\gls{cbn} interface through C-B bonded pair is predicted to be \SI{11.0 \pm 0.1}{\giga\watt\per\meter\squared\per\kelvin}, establishing a new upper bound among all existing theoretical predictions and experimental measurements. Our results demonstrate that the \gls{itc} is primarily governed by unique interface phonon modes, namely extended modes and localized modes, which are completely different from bulk-like phonon modes. Furthermore, when acoustic phonon mode matching between diamond and c-BN is disrupted by atomic diffusion, the contribution of extended modes to \gls{itc} is significantly diminished, and the presence of a diffuse interface severely impedes heat transfer. The highly localized modes with bridging effects also vanish under such disorder. Our work provides atomistic insights into the ideal limit of thermal transport across semiconductor interfaces, as well as the emergence and evolution of interfacial phonon modes that govern this limit, which may offer valuable guidance for related thermal management applications.

\section{methods}
\subsection{DFT calculations}
\label{section:dft}
All \gls{dft} calculations were performed using the Perdew-Burke-Ernzerhof functional within the generalized gradient approximation \cite{blochl1994projector,perdew1996generalized}, as implemented in the \textsc{vasp} package \cite{Kresse1996PRB,Kresse1999PRB}. A plane-wave basis set with an energy cutoff of \SI{600}{\electronvolt} was employed, ensuring energy convergence within \SI{1e-6}{\electronvolt} in the electronic self-consistent loop. The Brillouin zone was sampled using a $\Gamma$-centered $k$-point grid with a density of \SI{0.25}{\per\angstrom}, and Gaussian smearing with a width of \SI{0.05}{\electronvolt} was applied. 

To calculate the phonon dispersion of bulk diamond and \gls{cbn}, as well as their ideal interface, the structural optimizations were first performed  with an atomic force convergence threshold of \SI{1e-3}{\electronvolt\per\angstrom}. After that, the second-order interatomic force constants for optimized structures were calculated using density functional perturbation theory, as implemented in the \textsc{phonopy} package \cite{togo2015first}. For bulk diamond and \gls{cbn}, \numproduct{5x5x5} supercells were employed, whereas the $\left[111\right]$ diamond/\gls{cbn} interface was constructed by combining \numproduct{3x3x4} supercells of the primitive cells of both diamond and \gls{cbn}, with C-B bonded pair at the interface.

In calculating the third-order interatomic force constants, the structures are generated by the random displacement method as implemented in the  the Thirdorder script \cite{li2014Sheng} and the \numproduct{4x4x4} supercell is used for two bulk systems. To obtain accurate \gls{kappa}, we consider the interactions up to the seventh nearest neighbors. Based on second- and third-order interatomic force constants, the
\gls{kappa} can be obtained by iteratively solving the linearized \gls{bte} using \numproduct{5x5x5} supercells as implemented by ShengBTE package \cite{li2014Sheng}. The broadening factor was set to 0.1. A final q-points mesh of \numproduct{31x31x31} was found to yield well-converged \gls{kappa} values.

\subsection{NEP training}
\label{section:nep}
Our reference structures include bulk diamond, \gls{cbn}, and diamond/\gls{cbn} heterostructures with both ideal flat and rough interfaces (see \autoref{fig:nep} (a-d)), generated through \gls{aimd} sampling and perturbation. \gls{aimd} simulations were conducted at temperatures ranging from \SI{10}{\kelvin} to \SI{1000}{\kelvin} over 10,000 steps, with a time step of \SI{1}{\femto\second}. Perturbations were introduced by applying random cell deformations from -3\% to 3\% and atomic displacements within \SI{0.1}{\angstrom}.

For bulk diamond and \gls{cbn}, \numproduct{4x4x4} supercells of the primitive cell, each containing 128 atoms, were constructed. We sampled 200 structures from \gls{aimd} and 50 structures from perturbation for both diamond and \gls{cbn}, resulting in a total of 500 reference structures.

Mimicking the growth of \gls{cbn} on diamond in the experiment, the diamond/\gls{cbn} heterostructures with ideal interfaces along the  $\left[111 \right]$ interface was established. Compared with the $\left[100 \right]$ direction, its interface binding energy is higher, ensuring enhanced thermodynamic stability \cite{zhao2019electronic}. See \autoref{fig:nep}(c), we constructed a system composed of \numproduct{3x3x4} supercells of both diamond and \gls{cbn} primitive cells, containing a total of 144 atoms. We obtained 250 reference structures, including 200 from \gls{aimd} sampling and 50 from perturbation.

For diamond/\gls{cbn} heterostructures with rough interfaces, the interface model containing mutual diffusion layers is mainly constructed by randomly shuffle atoms, where different diffusion ratios represent interfaces of different roughness
Here, two configurations were considered to account for atomic diffusion effects: (i) we randomly replaced boron or nitrogen atoms with carbon atoms in \numproduct{4x4x4} supercells of the primitive cell of \gls{cbn} in increments of 10\%; (ii) we constructed a 20-layer heterostructure comprising \numproduct{2x2x10} supercells of the primitive cell for both diamond and \gls{cbn}, containing 160 atoms in total, where in the central eight layers, carbon and boron or nitrogen atoms were randomly mixed, again in 10\% incremental steps. All obtained structures, after atomic replacements or exchanges, were further subjected to perturbation. We generated 180 reference structures, with 90 structures for each configuration.

In total, we obtained 930 reference structures and performed single-point \gls{dft} calculations (see \autoref{section:dft} for details) on these structures to obtain the corresponding energy, forces, and virial data for subsequent \gls{nep} training. The complete reference dataset was randomly divided into a training set with 740 structures and a test set with 190 structures.

Using the obtained training and test datasets as input, we employed the NEP3 framework \cite{fan2022gpumd}, implemented in the \textsc{gpumd} package (verison v3.5) \cite{fan2017efficient}, to train a unified machine-learned \gls{nep} for diamond, \gls{cbn}, and their heterostructures. \gls{nep} employs a feedforward \gls{nn} to represent atomic site energy as a function of a descriptor vector containing radial and angular components, while \gls{snes} \cite{Schaul2011} optimizes the parameters to minimize the \glspl{rmse} of energy, forces, and virial against the training dataset. The cutoff radii for both radial and angular descriptor terms were set to \SI{4.5}{\angstrom}. A feedforward \gls{nn} with a hidden layer of 50 neurons was used. The separable natural evolution strategy algorithm was applied with a population size of 50, and a total number of $5\times 10^5$ generations was used to achieve convergence of the total loss function. The weights of energy, force, and virial \glspl{rmse} in the loss function were set to 1.0, 1.0, and 0.1, respectively. For more details of the NEP approach, we refer to the literature \cite{fan2022gpumd}.

\subsection{NEMD simulations}
\label{section:nemd}
All \gls{md} simulations were performed using the \textsc{gpumd} package (version v3.5) \cite{fan2017efficient}. The \gls{nemd} approach was employed to investigate the thermal transport properties of diamond/\gls{cbn} interfaces by establishing a non-equilibrium steady state with a constant heat flux using two local thermostats at different temperatures. A rectangular simulation box with dimensions of \SI{4.4}{\nano\meter}$\times$\SI{3.8}{\nano\meter}$\times$\SI{22.8}{\nano\meter}, containing 66,000 atoms, was used for all diamond/\gls{cbn} heterostructures. This system size was tested and confirmed to be sufficiently large to obtain convergent \gls{itc}. Periodic boundary conditions were applied in all three spatial directions, and the heat flux direction was set from the diamond side to the \gls{cbn} side (along the $z$-direction in \autoref{fig:nemd}(a)).

To ensure unidirectional thermal transport, atoms in the outermost five layers of both the diamond and \gls{cbn} regions were fixed. Adjacent to these fixed layers, five-layer heat source and heat sink regions were introduced to generate the heat flux. The transport region, located between the heat source and sink, consisted of 90 atomic layers: 45 layers of diamond on the left and 45 layers of \gls{cbn} on the right, with each layer containing 600 atoms. Moving outward from the center of the system, the layers on the left diamond side were labeled as $L1$, $L2$, $L3$, etc., while those on the right \gls{cbn} side were labeled as $R1$, $R2$, $R3$, etc.

For the \gls{nemd} simulations, the diamond/\gls{cbn} heterostructure was first relaxed for \SI{100}{\pico\second} at \SI{300}{\kelvin} using the Berendsen thermostat \cite{berendsen1984molecular} under the NVT ensemble. After relaxation, Langevin thermostats \cite{bussi2007accurate} were applied to the heat source and sink regions, maintaining temperatures of \SI{325}{\kelvin} and \SI{275}{\kelvin}, respectively. The entire system reached a steady state within \SI{2}{\nano\second}, after which temperature and energy profiles were sampled over the last \SI{1}{\nano\second}.

The \gls{itc} is defined in terms of the temperature drop $\Delta T$ across an interface as:  
\begin{equation}
\label{equation:G}
G=\frac{\langle J \rangle}{A \Delta T},
\end{equation}  
where $\left\langle J \right\rangle$ represents the average energy transfer rate along the temperature gradient direction, and $A$ is the cross-sectional area perpendicular to the transport direction. To fairly compare the \gls{itc} of ideal and rough interfaces, $\Delta T$ is calculated as the temperature difference between $L6$ and $R6$, where atomic diffusion is confined between these two layers (see \autoref{fig:nemd}(b)). For each case, three independent simulations were conducted to determine the average value as the predicted \gls{itc}, and the corresponding standard error was also calculated.

\subsection{Spectral heat current decomposition}
\label{section:shc}
To obtain the contribution of phonon modes with different frequencies, one can calculate spectrally decomposed thermal conductance $G(\omega)$ in the \gls{nemd} approach\cite{fan2019homogeneous}:
\begin{equation}
G = \int_0^{\infty} \frac{d\omega}{2\pi} G(\omega),
\end{equation}
where
\begin{equation}
\label{equation:Gw}
G(\omega) = \frac{2}{V\Delta T}\int_{-\infty}^{+\infty}e^{i\omega t} K(t) dt.
\end{equation}
Here, $K(t)$ is the virial-velocity-time correlation function \cite{gabourie2021spectral} in the transport direction. The full vector of the virial-velocity correlation function is defined as
\begin{equation}
    \bm{K}(t) = \sum_i \langle \mathbf{W}_i(0) \cdot \bm{v}_i(t) \rangle,
\end{equation}
where $\mathbf{W}_i$ is the virial tensor and $\mathbf{v}_i$ is the velocity vector of atom $i$.

Similar to thermal conductance, the \gls{kappa} can also be spectrally decomposed in the \gls{hnemd} approach:
\begin{equation}
\kappa= \int_{0}^{\infty}\frac{d\omega}{2\pi}\kappa(\omega);  
\end{equation}
\begin{equation}
\label{equation:Kw}
\kappa(\omega) = \frac{2}{VTF_{\rm e}}\int_{-\infty}^{+\infty}e^{i\omega t} K(t) dt,
\end{equation}
where $F_{\rm e}$ is the driving force parameters in the \gls{hnemd} method.

\begin{acknowledgments}
The authors thank Mr. Ting Liang and Dr. Ke Xu for their help in validating the NEP model.
This work was supported by the Key-Area Research and Development Program of Guangdong Province (Grant 2020B010169002), the Guangdong Special Support Program (Grant No.2021TQ06C953), and the Science and Technology Planning Project of Shenzhen Municipality (Grant  No. GXWD20220811164433002).
X. Wu is the JSPS Postdoctoral Fellow for Research in Japan (No. P24058). 
P. Ying is supported by the Israel Academy of Sciences and Humanities \& Council for Higher Education Excellence Fellowship Program for International Postdoctoral Researchers. 
\end{acknowledgments}

\noindent{\textbf{Conflict of Interest}}

The authors have no conflicts to disclose.

\ 
\

\noindent{\textbf{Data availability}}

The source code and documentation for  \textsc{gpumd} are available
at \url{https://github.com/brucefan1983/GPUMD} and \url{https://gpumd.org}, respectively.
The inputs and outputs related to the NEP model training are freely available at the Gitlab repository \url{https://gitlab.com/brucefan1983/nep-data}.

\end{document}